\documentclass[lettersize,journal]{IEEEtran}
\usepackage{amsmath,amsfonts}
\usepackage{algorithmic}
\usepackage{algorithm}
\usepackage{array}
\usepackage[caption=false,font=normalsize,labelfont=sf,textfont=sf]{subfig}
\usepackage{textcomp}
\usepackage{stfloats}
\usepackage{url}
\usepackage{verbatim}
\usepackage{graphicx}
\usepackage{xcolor}
\usepackage[sorting=none,style=ieee,backend=bibtex]{biblatex}
\usepackage{hyperref}
\hypersetup{hidelinks}

\usepackage{booktabs}
\usepackage{multicol}
\usepackage{multirow}

\begin{document}

\title{AeroGPT: Leveraging Large-Scale Audio Model for Aero-Engine Bearing Fault Diagnosis}

\author{Jiale Liu, Dandan Peng, Huan Wang*, Chenyu Liu, Yan-Fu Li,~\IEEEmembership{Senior Member,~IEEE}, \\ and Min Xie,~\IEEEmembership{Fellow,~IEEE}
        % <-this % stops a space
\thanks{Jiale Liu is with the School of Physics and Astronomy, The University of Edinburgh, Edinburgh, UK, and the Glasgow College, University of Electronic Science and Technology of China, Chengdu, China.}
\thanks{Dandan Peng is with the Department of Electrical and Electronic Engineering, The Hong Kong Polytechnic University, Hong Kong, China.}
\thanks{Huan Wang is with the Department of Systems Engineering, City University of Hong Kong, Hong Kong, China.}
\thanks{Chenyu Liu is with the School of Mechanical Engineering, Northwestern Polytechnical University, Xi'an, China.}
\thanks{Yan-Fu Li is with the Department of Industrial Engineering, Tsinghua University, Beijing, China.}
\thanks{Min Xie is with the Department of Systems Engineering, City University of Hong Kong, Hong Kong, China and the City University of Hong Kong Shenzhen Research Institute, Shenzhen, China.}
\thanks{This work is supported by National Natural Science Foundation of China (72371215, 72032005), Research Grant Council of Hong Kong (11201023, 11202224, Project No. CityU JRFS2526-1S09), and the Beijing Municipal Natural Science Foundation-Rail Transit Joint Research Program (L231020). Corresponding author: Huan Wang, wh.2021@tsinghua.org.cn.}% <-this % stops a space
\thanks{This work has been submitted to the IEEE for possible publication. Copyright may be transferred without notice, after which this version may no longer be accessible.}
}

\markboth{Paper Accepted by IEEE Transactions on Cybernetics}%
{}


\maketitle

\begin{abstract}
Aerospace engines, as critical components in aviation and aerospace industries, require continuous and accurate fault diagnosis to ensure operational safety and prevent catastrophic failures. {While deep learning techniques have been extensively studied in this context, they typically output logits or confidence scores, necessitating post-processing to obtain actionable insights. Furthermore, the potential of large-scale audio models for this task remains largely untapped.} To address these limitations, this paper proposes AeroGPT, a novel framework that transfers knowledge from the general audio domain to aero-engine bearing fault diagnosis. {AeroGPT leverages a large-scale audio model and incorporates Vibration Signal Alignment (VSA) to adapt general audio knowledge to domain-specific vibration patterns, along with Generative Fault Classification (GFC) to directly generate interpretable fault labels.} {This approach eliminates the need for label post-processing and supports interactive, interpretable, and actionable fault diagnosis, thereby enhancing industrial applicability. Through comprehensive experimental validation on two aero-engine bearing datasets, AeroGPT achieves 98.94\% accuracy on the DIRG dataset and 100\% accuracy on the HIT bearing dataset, outperforming representative deep learning approaches. Qualitative analysis and further discussion also demonstrate its potential for interactive diagnosis and real-world deployment, highlighting the promise of large-scale audio models to advance fault diagnosis in aerospace applications.}
\end{abstract}


\section{Introduction}
\IEEEPARstart{A}{erospace} engines serve as the cornerstone of modern aviation and aerospace industries, powering everything from commercial aircraft to space exploration vehicles \cite{chen2022casebased, lin2023novel}. Their reliable operation is paramount to aviation safety, mission success, and the prevention of catastrophic failures that could result in significant human and economic losses. Within these sophisticated propulsion systems, aero-engine bearings are particularly critical components, operating under extreme conditions of high speeds, temperature variations, and significant mechanical stress \cite{huang2023novel, liu2024braininspired, zhang2020aeroengine}. The failure of these bearings can lead to severe engine malfunctions, making them a primary focus for condition monitoring and maintenance protocols \cite{ding_multitask_2024, lin2023novel}. Fault diagnosis of aero-engine bearings has consequently emerged as an essential discipline, offering methodologies to detect, identify, and predict potential failures before they manifest as catastrophic events. Early and accurate bearing fault detection can significantly extend engine lifespan, optimize maintenance schedules, and ensure operational safety \cite{wang2024highaccuracy}. 

Traditionally, bearing fault diagnosis approaches rely on a two-step process of feature extraction followed by classification \cite{peng_automatic_2021}. For instance, Medina \textit{et al.} \cite{Medina2020Gear} proposed an approach using Poincare plots for feature extraction, with subsequent classification performed by a multi-class Support Vector Machine (SVM). However, such approaches are limited by their dependence on domain-specific feature engineering and conventional machine learning classifiers, which may fail to fully capture the complex dynamics of aero-engine bearing systems. While the subsequent rise of deep learning methods has addressed the challenge of manual feature engineering, they introduce a different limitation. These data-driven methods typically output logits or confidence scores, necessitating post-processing to derive actionable insights, whereas industrial applications require efficient and interpretable solutions.

Recently, the emergence and development of large language models (LLMs) and large multimodal models (LMMs) offer transformative advantages in this regard. These models have demonstrated remarkable capabilities in generating human-like text, understanding complex queries, and providing contextually relevant responses. With interactivity, they can incorporate contextual inputs or user queries to provide tailored, situation-specific diagnoses. Furthermore, they can be designed to directly output decisions or actionable recommendations, eliminating the need for complex post-processing. More critically, LLMs shift the paradigm from low-information-density fault categories to high-information-density analysis, {delivering detailed insights beyond defect categories, thereby enhancing both interpretability and practical utility in industrial applications.}

Building upon these observations, we identify a critical oversight in current approaches: whether utilizing the text modality of LLMs or extending to the visual modality of LMMs, these methods fail to recognize a fundamental characteristic of bearing vibration signals, which is their intrinsic similarity to acoustic phenomena. {Bearing vibration signals, as mechanical oscillations propagating through a medium, manifest as time-varying waveforms with frequency, amplitude, and temporal patterns that closely resemble those of sound waves.} This acoustic likeness is evident in their shared spectral properties, such as harmonic peaks and resonance frequencies, which are routinely analyzed in audio processing but underutilized in fault diagnosis. {This observation reveals a significant methodological gap in current research: traditional deep learning methods implement a ``signal $\rightarrow$ logits'' pipeline, outputting probability distributions that require post-processing, while recent works leveraging large language models have adopted indirect approaches like converting vibration signals into time-frequency images for vision-language models, or textualizing signal features before feeding them to LLMs. However, no prior work has established a direct ``signal $\rightarrow$ language'' paradigm that preserves the native waveform structure.} Furthermore, despite the critical importance of aerospace propulsion systems and the severe consequences of their failure, large-scale models specifically designed for aero-engine bearing fault diagnosis have yet to be developed. The absence of such specialized models represents a significant gap in the current research landscape and limits the effectiveness of predictive maintenance in aerospace applications.

{Motivated by the above challenges, this paper proposes AeroGPT, a novel framework that aligns general-domain audio knowledge with aero-engine bearing vibration patterns for fault diagnosis in aerospace engines. The core novelty of AeroGPT lies in modality-aligned transfer and task reformulation, which make large-scale audio modeling directly applicable to aero-engine bearing fault diagnosis by treating vibration signals as audio-like waveforms and leveraging the acoustic representation capability of audio foundation models. This design is motivated by the fact that bearing vibration signals are fundamentally mechanical oscillations with temporal–spectral characteristics that closely resemble acoustic phenomena. By exploiting this inherent similarity, AeroGPT can directly utilize rich pre-trained knowledge from audio foundation models without the information loss introduced by time–frequency image conversion or feature textualization. As a large-scale audio model-based approach, AeroGPT addresses the limitations of current methods through two key innovations: Vibration Signal Alignment (VSA), which adapts general audio knowledge to the specific characteristics of aero-engine bearing vibrations, and Generative Fault Classification (GFC), which preserves the generative capability of the underlying LLM to directly output interpretable fault labels, eliminating the need for post-processing. To enable efficient adaptation to the aero-engine bearing domain, we employ Low-Rank Adaptation (LoRA), a parameter-efficient fine-tuning method that freezes most pre-trained parameters and learns only low-rank updates, substantially reducing computational and memory costs while maintaining strong performance. By leveraging the generative capabilities of large-scale audio models, AeroGPT enables interactive, interpretable, and actionable fault diagnosis, enhancing its practical utility in aerospace applications. Experiments on two aero-engine bearing datasets demonstrate that AeroGPT achieves 98.94\% accuracy on the DIRG dataset and 100\% accuracy on the HIT bearing dataset, outperforming traditional deep learning approaches. Additional qualitative analysis and discussion further demonstrate the applicability and potential of AeroGPT to advance fault diagnosis in safety-critical aerospace systems.}

The main contributions of this paper can be summarized as follows:
\begin{enumerate}
    \item {To address the challenge that existing fault diagnosis methods rely on indirect signal representations and output non-interpretable logits, this paper is the first to adapt a large-scale audio model for industrial fault diagnosis, proposing a modality-aligned paradigm that directly processes vibration signals as audio-like waveforms and generates human-readable diagnostic outputs.}
    \item {To bridge the domain gap between general audio knowledge and mechanical vibration patterns, this paper proposes Vibration Signal Alignment (VSA), an intermediate adaptation objective that explicitly learns vibration-language grounding through vibration-text pairs, enabling effective knowledge transfer to the aero-engine bearing domain.}
    \item {To meet industrial requirements for interpretability and actionability, this paper proposes Generative Fault Classification (GFC), which leverages the inherent generative capabilities of LLMs to directly output structured, human-readable fault labels and support interactive follow-up analysis, eliminating the need for post-processing pipelines.}
    \item {By integrating the above components, this paper proposes AeroGPT, a unified framework that achieves state-of-the-art performance on two aero-engine bearing datasets (98.94\% on DIRG and 100\% on HIT), demonstrating the effectiveness of the vibration-as-audio paradigm and the potential of large-scale audio models for aerospace fault diagnosis.}
\end{enumerate}

The remainder of this paper is structured as follows: Section II reviews the related work on fault diagnosis; Section III details the methodology of the proposed AeroGPT framework, covering its system architecture, the audio knowledge acquisition process, LoRA-based domain knowledge adaptation, and the core mechanisms of Vibration Signal Alignment (VSA) and Generative Fault Classification (GFC); Section IV presents our experimental validation, including comprehensive evaluations on the DIRG and HIT bearing datasets; {Section V provides further discussion on the computational considerations and safe integration strategies}; finally, Section VI concludes the paper by summarizing our findings and outlining directions for future research.

{\section{Related Work}}

{The field of bearing fault diagnosis has evolved significantly, primarily driven by advances in machine learning techniques. This section reviews the related literature by categorizing prior work into two paradigms: the well-established discriminative approaches powered by deep learning, and the emerging generative approaches leveraging large-scale models.}

{\subsection{Discriminative Approaches with Deep Learning}}
{The first paradigm, which has been the mainstream of research for the past few years, treats fault diagnosis as a discriminative classification task. The goal of these methods is to learn a mapping from raw signal inputs to a set of predefined fault labels by identifying discriminative features. This paradigm has been significantly empowered by data-driven methods, particularly deep learning approaches, which enable end-to-end learning directly from raw vibration signals \cite{chen_multi-source_2023, xu_deep_2024}. These methods automatically extract hierarchical features, eliminating the need for the manual feature engineering that traditional approaches require.}

{Convolutional Neural Networks (CNNs) have been widely adopted for this purpose, as they excel at capturing local patterns and spatial hierarchies in data. Wang \textit{et al.} \cite{wang2020understanding} proposed a one-dimensional CNN with an attention mechanism tailored for bearing fault enhancement and classification, establishing a new state-of-the-art performance on the wheelset bearing dataset. Apart from CNNs, Recurrent Neural Networks, especially Long Short-Term Memory (LSTM) networks, have also gained traction. LSTMs are adept at capturing temporal dependencies in sequential data, making them suitable for modeling the time-varying nature of vibration signals. Song \textit{et al.} \cite{song2024optimized} developed a CNN-BiLSTM network that combines CNN's spatial feature extraction capabilities with BiLSTM's temporal modeling, demonstrating robust performance even with limited samples.}

{More recently, Transformer-based architectures \cite{vaswani2017attention}, leveraging self-attention mechanisms to capture long-range dependencies and contextual information, have also emerged as powerful tools. For instance, Li \textit{et al.} \cite{li2024dconformer} proposed Dconformer, a novel CNN-Transformer network that combines joint-learning denoising with a multi-branch cross-cascaded architecture to extract both local and global features from noisy vibration signals. Xiang \textit{et al.} \cite{xiang2025frequency} introduced a frequency channel-attention based Vision Transformer method, which integrates frequency domain characteristics with self-attention mechanisms. To address the high computational cost of Transformers, Guo \textit{et al.} \cite{guo_effective_2025} developed SPCFormer, a lightweight Transformer variant with selective patches and channel modules designed for efficient fault diagnosis in resource-constrained systems like quadrotor helicopters.}

{Beyond these architectures, researchers have explored other neural paradigms to address specific industrial challenges. For instance, Autoencoders (AEs) are widely used for unsupervised feature learning \cite{he_masked_2021, guo_multivariate_2025}. Guo \textit{et al.} \cite{guo_multivariate_2025} developed a multivariate fusion covariance matrix network (MFCMN), which feeds specially constructed covariance matrices into a standard autoencoder to effectively handle multi-channel signals with limited labeled samples. In another innovative direction, Graph Neural Networks (GNNs) have been introduced to utilize inter-sensor dependencies \cite{li_emerging_2022, li_filter-informed_2024}. Li \textit{et al.} \cite{li_filter-informed_2024} designed a spectral graph wavelet network (SGWN) capable of multiscale feature extraction while mitigating over-smoothing. Brain-inspired architectures like Spiking Neural Networks (SNNs) have also been explored, with Xu \textit{et al.} \cite{xu_deep_2024} proposing a deep spiking residual shrinkage network (DSRSN) that achieves high accuracy and efficiency in noisy environments.}

{Alongside these architectural innovations, significant research has focused on learning strategies to enhance model adaptability and reduce data dependency. Prominent examples include transfer learning \cite{li_perspective_2022, qin_deep_2023}, meta-learning \cite{ren_meta-learning_2024}, few-shot learning \cite{wang_few-shot_2023, ren_novel_2020}, and unsupervised learning \cite{wang2025multilabel}. However, the aforementioned methods are fundamentally discriminative, outputting logits that require post-processing, which is a limitation that our generative approach aims to overcome.}

{\subsection{Generative Approaches with Large-Scale Models}}
{A new paradigm is emerging with the advent of large-scale foundation models, which reformulates fault diagnosis as a generative task. Instead of merely classifying faults, these methods aim to generate rich, human-readable outputs, enabling interactive and interpretable diagnostics. The application of such generative models to industrial fault diagnosis remains in its early stages, but several noteworthy research directions are taking shape.}

{One major research thrust focuses on building comprehensive Industrial Foundation Models (IFMs) and enhancing their reasoning capabilities with structured knowledge. For instance, Ren \textit{et al.} \cite{ren_industrial_2025} proposed a system architecture for a general-purpose IFM, designed to handle diverse industrial modalities and tasks throughout a product's lifecycle. To bolster the reliability of LLMs in complex industrial settings, other researchers have turned to knowledge graphs (KGs). Zhuang \textit{et al.} \cite{zhuang_large_2025} integrated a time-frequency KG with a large model to improve fault semantic capture from multimodal data. Similarly, Zhou \textit{et al.} \cite{zhou_causalkgpt_2024} developed CausalKGPT, a framework that enhances an LLM with a causal knowledge graph to perform cause analysis for quality problems in aerospace manufacturing. Nie \textit{et al.} \cite{nie_industrial_2026} further integrated LLMs with KGs and Retrieval-Augmented Generation (RAG) for fault reasoning and maintenance recommendations in CNC machine tools. These approaches aim to inject LLMs with deep, structured domain knowledge, making them more trustworthy and capable of complex reasoning.}

{A second, more task-specific direction involves developing multimodal diagnostic models that fuse different representations of machine health data. BearingFM \cite{lai2024bearingfm} established a framework for training large-scale models in this domain, utilizing domain knowledge-based data augmentation and contrastive learning to extract features from unlabeled vibration signals, achieving high accuracy with minimal labeled data. Tao \textit{et al.} \cite{tao2025llmbased} leveraged the text modality by quantitatively selecting features of vibration signals to textualize the time-series data and utilized LoRA for fine-tuning LLMs, demonstrating how language models can interpret bearing conditions. Taking a different approach, FaultGPT \cite{chen2025faultgpt} exploited the vision modality by extracting features from vibration time-frequency images, which were then paired with textual instructions to generate detailed diagnostic reports. Wang \textit{et al.} \cite{wang_diagllm_2025} proposed DiagLLM, which incorporates multimodal signal representations and expert knowledge to enable explainable bearing fault diagnosis. More recently, Li \textit{et al.} \cite{li_fd-mvllm_2025} proposed FD-MVLLM, which combines raw time-series signals with their time-frequency image representations, reprogramming both modalities into a format digestible by an LLM to improve diagnostic accuracy. Zhang \textit{et al.} \cite{zhang_fault_2026} further explored discretized signal representations and lightweight adapters to bridge continuous vibration signals with the discrete token inputs of LLMs.}

{These pioneering works validate the potential of large-scale models in various industrial fields but also highlight the need for a more direct and modality-aligned approach. While IFMs and KG-enhanced models offer powerful reasoning, they cannot directly use the domain knowledge to perform signal-based fault diagnosis tasks. {Meanwhile, existing multimodal diagnostic models typically rely on indirect signal representations like images, textualized features, or discretized tokens.} Our work contributes to this emerging paradigm by being the first to treat vibration signals as audio, thereby addressing a key gap in the current state of the art.}

\section{Methodology}
\subsection{Overview of AeroGPT}
\begin{figure*}[htbp]
    \centering
    \includegraphics[width=0.95\linewidth]{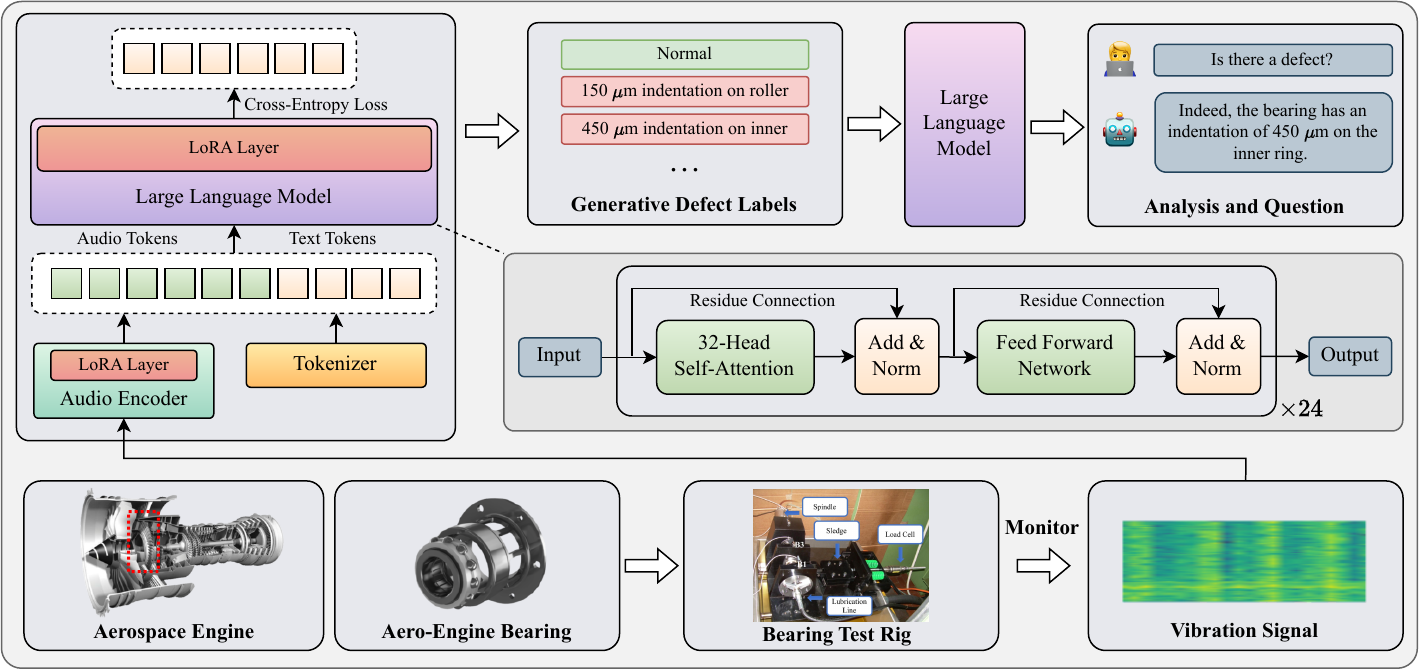}
    \caption{Overall framework of AeroGPT and its practical application to aero-engine bearing fault diagnosis.}
    \label{fig:integration}
\end{figure*}

\begin{figure*}[htbp]
    \centering
    \includegraphics[width=\linewidth]{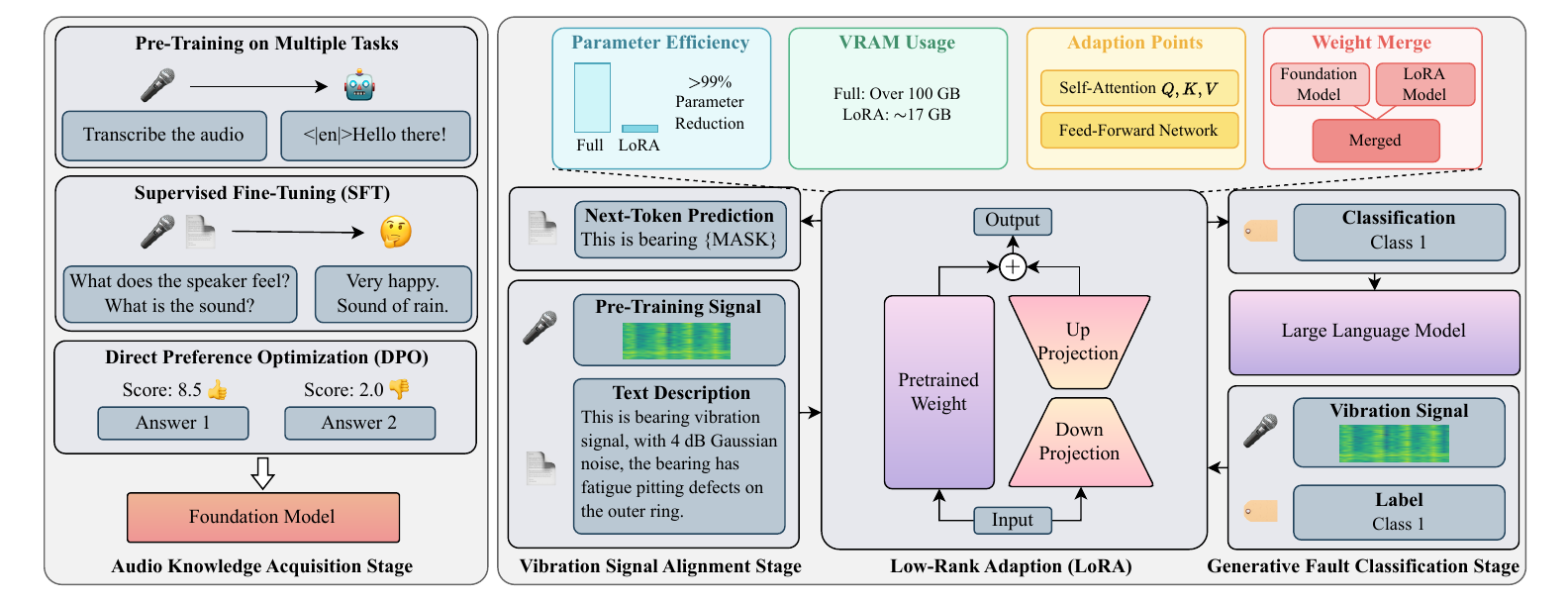}
    \caption{{Technical components of the AeroGPT methodology. The framework is initialized with a foundation model, followed by Vibration Signal Alignment Stage to adapt general audio knowledge to domain-specific vibration patterns and Generative Fault Classification Stage to output interpretable fault labels.}}
    \label{fig:train}
\end{figure*}

For aero-engine bearing fault diagnosis, this paper proposes AeroGPT, a novel framework leveraging large-scale audio models, and the overall pipeline is shown in \autoref{fig:integration}. {This framework explores the application of large-scale audio models to aerospace bearing fault diagnosis for the first time}, with the aim of generating interpretable and actionable diagnostic outputs directly from vibration signals. AeroGPT transforms the conventional classification-based fault diagnosis approach into an interactive, generative process capable of providing detailed insights beyond simple fault categorization.

As shown in \autoref{fig:integration}, the vibration signals of the aero-engine bearings are collected from the bearing test rig and then input into an audio encoder equipped with a LoRA layer, which effectively converts the vibration patterns into audio tokens for LLM understanding. The input representation process can be formalized as:
\begin{equation}
\mathcal{I}_{\text{AeroGPT}} = \{\mathcal{E}_{\text{audio}}(\mathbf{v}_{\text{proc}}), \mathcal{E}_{\text{text}}(\mathbf{p}), \mathcal{E}_{\text{text}}(\mathbf{c})\}\end{equation}
where $\mathcal{E}_{\text{audio}}$ and $\mathcal{E}_{\text{text}}$ are the audio and text encoders respectively, $\mathbf{v}_{\text{proc}}$ is the processed vibration signal, $\mathbf{p}$ is the prompt template, and $\mathbf{c}$ represents optional context information. The audio tokens and text tokens are then concatenated together as input to the LLM, which then processes these tokens and generates a fault label as the output. Leveraging the generative capabilities of the LLM, analysis and follow-up questions can also be conducted interactively, allowing for a more comprehensive understanding of the fault condition.

AeroGPT incorporates two key mechanisms that address fundamental challenges in utilizing large-scale audio models for fault diagnosis, as illustrated in \autoref{fig:train}. The first component, Vibration Signal Alignment (VSA), bridges the gap between general audio knowledge and the specific characteristics of aero-engine bearing vibrations. VSA processes paired inputs of vibration signals and corresponding textual descriptions, enabling the model to establish meaningful connections between acoustic patterns and their diagnostic interpretations. {The second component}, Generative Fault Classification (GFC), leverages the inherent capabilities of LLMs to directly generate diagnostic outputs in natural language rather than numerical logits that require post-processing. This approach provides actionable insights for maintenance personnel and supports interactive diagnostic sessions where follow-up questions can be addressed, demonstrating practical utility of the system in aerospace applications. Detailed descriptions of the components are provided in the following sections.

\subsection{Audio Knowledge Acquisition}
The acoustic understanding capabilities in AeroGPT are developed through a multi-stage process, as illustrated in the leftmost part of \autoref{fig:train}. The initial stage involves exposure to diverse audio-related tasks such as transcription, enabling the model to develop fundamental representations of acoustic patterns and their semantic interpretations. Subsequently, supervised fine-tuning (SFT) enhances the model's analytical capabilities through higher-level audio understanding tasks, cultivating the ability to extract contextual information and identify subtle acoustic features.

The final refinement stage implements Direct Preference Optimization (DPO), where alternative responses to the same audio input are evaluated through a comparative scoring mechanism. The optimization objective is formulated as:
\begin{equation}
    \mathcal{L}_{DPO} = -\mathbb{E} \Biggl[ \log \sigma \Biggl( \beta \log \frac{\mathcal{P}_\theta(y_w|x)}{\mathcal{P}_{ref}(y_w|x)} - \beta \log \frac{\mathcal{P}_\theta(y_l|x)}{\mathcal{P}_{ref}(y_l|x)} \Biggr) \Biggr],
\end{equation}
where $\mathcal{P}_\theta$ represents the system being optimized, $\mathcal{P}_{ref}$ is a reference system, $y_w$ and $y_l$ are the preferred and less preferred responses respectively, $\sigma$ is the sigmoid function, and $\beta$ is a hyperparameter controlling optimization strength. This multi-stage process produces a large-scale audio foundation model with sophisticated acoustic pattern recognition capabilities, providing the foundation for interpreting bearing vibration signals as acoustic phenomena.

\subsection{LoRA-based Domain Knowledge Adaptation}

To effectively transfer general audio knowledge to aero-engine bearing fault diagnosis, AeroGPT employs a domain adaptation approach based on Low-Rank Adaptation (LoRA) techniques. This adaptation enables efficient knowledge transfer while maintaining computational feasibility even with limited resources.

The adaptation process operates on the principle that domain-specific knowledge can be effectively integrated through low-rank decomposition of weight updates. Rather than modifying the entire weight matrices of the large-scale model, which would require enormous computational resources, AeroGPT freezes the pre-trained weight matrices $\mathbf{W}_0 \in \mathbb{R}^{d \times k}$ and introduces trainable low-rank adaptations through:
\begin{equation}
\mathbf{h} = \mathbf{W}_0 \mathbf{x} + \frac{\alpha}{r} \mathbf{B}\mathbf{A} \mathbf{x}
\end{equation}
where $\mathbf{B} \in \mathbb{R}^{d \times r}$ and $\mathbf{A} \in \mathbb{R}^{r \times k}$ are low-rank matrices with rank $r \ll \min(d, k)$, $\mathbf{x}$ is the input, $\mathbf{h}$ is the output, and $\alpha$ is a scaling factor for training stability. This factorization significantly reduces trainable parameters from $d \times k$ to $r \times (d + k)$, achieving over 99\% parameter reduction and reducing VRAM usage from over {100 GB} for full fine-tuning to approximately {17 GB}.

In our framework, LoRA adaptations are applied to all linear layers within the audio encoder, aligner, and LLM components, including self-attention mechanisms (query, key, value projections) and feed-forward networks. {The LoRA rank $r$ is set to 16 and the scaling factor $\alpha$ is set to 32. Matrix $\mathbf{A}$ is initialized with random Gaussian values, while $\mathbf{B}$ is initialized with zeros to ensure $\Delta\mathbf{W} = \mathbf{B}\mathbf{A} = 0$ at training start.} Through this adaptation process, AeroGPT bridges the gap between general audio understanding and aero-engine bearing vibrations.

\subsection{Vibration Signal Alignment (VSA)}
The Vibration Signal Alignment (VSA) mechanism is designed to bridge the fundamental gap between general audio knowledge and the domain-specific characteristics of aero-engine bearing vibrations. {This is achieved through an alignment process that preserves the rich hierarchical representations inherent in large-scale audio models while adapting them to the specialized patterns of mechanical vibrations.}

The proper alignment requires establishing connections between vibration signals and their corresponding semantic interpretations. To facilitate this, we construct a dataset of vibration signals paired with detailed textual descriptions that articulate their diagnostic significance. Each description systematically captures critical fault-related information, including vibration source, noise characteristics, fault type, severity, location, and relevant operating conditions. This pairing strategy enables the model to develop a nuanced understanding of the relationship between acoustic features and their diagnostic implications. To enhance generalizability and prevent overfitting, the vibration data contains diverse bearing sources that differ from those used in our final aerospace validation tests.

The alignment process begins with precise signal preparation to ensure optimal encoding. Raw vibration signals $\mathbf{v}_{\text{raw}} \in \mathbb{R}^{T}$ undergo amplitude normalization through a statistical transformation:
\begin{equation}
\mathbf{v}_{\text{proc}} = \mathcal{F}_{\text{norm}}(\mathbf{v}_{\text{raw}}; \alpha, \beta) = \beta \cdot \frac{\mathbf{v}_{\text{raw}} - \mu_{\mathbf{v}}}{\sigma_{\mathbf{v}}} + \alpha
\end{equation}
where $\mu_{\mathbf{v}}$ and $\sigma_{\mathbf{v}}$ are the mean and standard deviation of the signal, and $\alpha, \beta$ are calibration hyperparameters optimized for the audio encoder's dynamic range. This normalization ensures consistent amplitude profiles across diverse operational conditions and sensor configurations, significantly enhancing the model's robustness to variations in signal acquisition parameters.

The normalized vibration signals are then transformed into audio embeddings through our domain-adapted encoder:
\begin{equation}
\mathbf{z}_{\text{audio}} = \mathcal{E}_{\text{audio}}(\mathbf{v}_{\text{proc}}; \theta_{\text{base}} + \Delta\theta_{\text{LoRA}})
\end{equation}
where $\mathcal{E}_{\text{audio}}$ represents the audio encoder, $\theta_{\text{base}}$ denotes the frozen parameters of the pre-trained model, and $\Delta\theta_{\text{LoRA}}$ encompasses the trainable LoRA parameters. This formulation enables selective adaptation of the encoder's parameters while preserving its foundational acoustic understanding. The encoder projects temporal vibration patterns into a high-dimensional embedding space $\mathbf{z}_{\text{audio}} \in \mathbb{R}^{L_a \times d}$, where $L_a$ represents the sequence length and $d$ is the embedding dimension.

In parallel, textual prompts and context information are processed through a text encoder to obtain corresponding embeddings:
\begin{equation}
\mathbf{z}_{\text{text}} = \mathcal{E}_{\text{text}}(\mathbf{p}, \mathbf{c}; \theta_{\text{text}}) \in \mathbb{R}^{L_t \times d}
\end{equation}
where $\mathbf{p}$ represents the instruction prompt, $\mathbf{c}$ denotes optional context, $\theta_{\text{text}}$ are the parameters of the text encoder, $L_t$ is the text sequence length, and $d$ is the embedding dimension shared with the audio encoder to facilitate cross-modal interactions.

These audio embeddings are subsequently tokenized to create a discrete representation compatible with the language model's input format:
\begin{equation}
\mathbf{T}_{\text{audio}} = \text{Tokenize}(\mathbf{z}_{\text{audio}}) = \{\mathbf{t}^{(a)}_1, \mathbf{t}^{(a)}_2, ..., \mathbf{t}^{(a)}_{L_a}\} \in \mathbb{R}^{L_a \times d}
\end{equation}
where $\mathbf{T}_{\text{audio}}$ represents the sequence of audio tokens of length $L_a$, each with dimension $d$, derived from the encoded vibration signal.

The cross-modal attention mechanism then establishes connections between the audio and text modalities, computing attention weights $\mathbf{A} \in \mathbb{R}^{L_t \times L_a}$ as:
\begin{equation}
\mathbf{A} = \text{softmax}\left(\frac{\mathbf{Q}\mathbf{K}^{\top}}{\sqrt{d_k}}\right) = \text{softmax}\left(\frac{\mathbf{z}_{\text{text}}\mathbf{W}_Q (\mathbf{z}_{\text{audio}}\mathbf{W}_K)^{\top}}{\sqrt{d_k}}\right)
\label{eq:cross_attn}
\end{equation}
where $\mathbf{W}_Q$ and $\mathbf{W}_K$ are learnable projection matrices, and $d_k$ is the dimensionality of the key vectors. This mechanism allows the model to selectively attend to relevant regions of the vibration signal when interpreting or generating diagnostic text, effectively establishing a vibration-to-language mapping that captures subtle fault-indicative patterns.

The training objective is to predict the next token in an autoregressive manner, with the main goal being to optimize the model so that it can predict subsequent tokens in the text description based on vibration signals and previous text. During training process, the cross-entropy loss is minimized:
\begin{equation}
\mathcal{L}_{\text{CE}} = -\frac{1}{|\mathcal{D}|} \sum_{(\mathbf{v}, \mathbf{y}) \in \mathcal{D}} \sum_{t=1}^{|\mathbf{y}|} \log p_{\theta}(y_t | \mathbf{z}_{\text{audio}}, y_{<t})
\end{equation}
where $\mathcal{D}$ represents the training dataset of vibration-text pairs, $\mathbf{y}$ denotes the target textual description, $y_t$ is the $t$-th token in the description, $y_{<t}$ represents all preceding tokens, and $p_{\theta}$ is the probability distribution over the vocabulary predicted by the model with parameters $\theta$.

This training paradigm enables the model to learn contextual relationships between vibration patterns and diagnostic interpretations in an end-to-end manner while preserving the linguistic capabilities of the underlying large language model, {allowing for the generation of interpretable diagnostic outputs rather than mere classification logits.} Our experimental results demonstrate that this alignment strategy is able to transfer the general audio knowledge to the specific domain of aero-engine bearing vibrations, enhancing the model's ability to recognize and interpret complex fault patterns.

{The VSA stage utilizes the Paderborn University bearing dataset \cite{lessmeier2016condition} for constructing vibration-text pairs. This dataset contains 32 bearing samples across three fault categories (healthy, inner ring damage, outer ring damage) with diverse damage mechanisms: 6 healthy bearings with different run-in periods, 12 artificially damaged bearings created through EDM machining, drilling, and electric engraving, and 14 real-damage bearings exhibiting fatigue pitting and plastic deformation from accelerated life tests. We use this dataset for VSA rather than benchmarking because while its coarse-grained labels with only inner/outer ring distinction are too simple for fine-grained evaluation, its diverse damage mechanisms are ideal for learning generalizable vibration-semantic associations.}

{Since publicly available bearing fault datasets provide only numerical class labels without natural language descriptions, we constructed a dataset of 36{,}053 vibration-text pairs using a semi-automatic method that combines expression templates with domain knowledge. Per-sample human authoring was avoided due to extreme effort and the risk of inconsistent phrasing in large-scale corpora. The design objective was to keep each description faithful to the physical state of the signal while enabling wording diversity.}

{For each sample, a small set of fields was obtained from labels and test-rig metadata, including fault type (healthy, inner race, outer race, rolling element), fault location, severity (e.g., damage size such as 150~$\mu$m or a qualitative level such as mild or moderate), operating speed (e.g., 6000~rpm), applied load (e.g., 1800~N), and a brief summary of signal characteristics. The characteristic field was linked to the fault type using common diagnostic patterns: for example, inner-race faults often present periodic high-frequency impact pulses, whereas healthy samples exhibit stationary background noise.}

{Natural language descriptions were then generated by filling several sentence templates with these fields. {The templates employ different sentence structures and phrase variants (e.g., ``inner race''/``inner ring'', ``impact pulses''/``impulsive bursts''). During generation, we randomly select a template and phrase variants, yielding descriptions that remain faithful to the underlying data while adding modest variety.} This process provides consistent supervision for VSA and improves robustness and generalization.}

{It is important to note that template-based generation has inherent limitations. The linguistic diversity is constrained by the number of templates and synonym substitutions, which may cause the model to learn ``template style'' rather than fully open-ended diagnostic language. However, we try to mitigate these concerns through employing multiple sentence structures and safe phrase variants to introduce linguistic variety, including descriptive content about signal characteristics linked to fault types based on established diagnostic patterns, and monitoring throughout the training process to avoid overfitting to the templates and forgetting the true semantic meaning.}

\subsection{Generative Fault Classification (GFC)}
{The Generative Fault Classification (GFC) represents a paradigm shift in fault diagnosis by leveraging the inherent generative capabilities of large language models to directly produce interpretable diagnostic outputs. Unlike conventional deep learning approaches that output numerical logits requiring post-processing, GFC enables AeroGPT to generate human-readable fault labels and explanations in natural language, significantly enhancing both interpretability and actionability for maintenance personnel.}

Traditional fault classification methodologies typically employ multi-class classifiers that output probability distributions across predefined fault categories, necessitating argmax operations and label mapping to determine the final classification. In contrast, GFC capitalizes on the autoregressive nature of LLMs to generate textual fault labels directly. This approach eliminates intermediate post-processing steps, streamlining the diagnostic pipeline while providing outputs in a format that is immediately comprehensible to maintenance technicians. The generation process follows:
\begin{equation}
\hat{y}_t = \underset{y \in \mathcal{V}}{\arg\max} \: p_{\theta}(y|\mathbf{z}_{\text{audio}}, \hat{y}_{<t})
\end{equation}
where $\hat{y}_t$ represents the predicted token at position $t$, $\mathcal{V}$ denotes the vocabulary space, $\mathbf{z}_{\text{audio}}$ is the encoded representation of the vibration signal, and $\hat{y}_{<t}$ comprises all previously generated tokens. This autoregressive generation continues until a complete fault diagnosis is produced.

For model optimization, the training objective is formulated as a cross-entropy loss function over the sequence of tokens constituting the fault label:
\begin{equation}
\mathcal{L}_{\text{GFC}} = -\frac{1}{|\mathcal{D}_{\text{fault}}|} \sum_{(\mathbf{v}, \mathbf{y}) \in \mathcal{D}_{\text{fault}}} \sum_{t=1}^{|\mathbf{y}|} \log p_{\theta}(y_t | \mathbf{z}_{\text{audio}}, y_{<t})
\end{equation}
where $\mathcal{D}_{\text{fault}}$ represents the fault diagnosis dataset comprising vibration signal-label pairs $(\mathbf{v}, \mathbf{y})$, and $p_{\theta}(y_t | \mathbf{z}_{\text{audio}}, y_{<t})$ denotes the probability of generating the correct token $y_t$ given the audio embedding and preceding tokens. During parameter optimization, we utilize LoRA to selectively adapt the model while preserving its foundational generative capabilities, which can be represented by:
\begin{equation}
\theta_{\text{LoRA}}^* = \underset{\theta_{\text{LoRA}}}{\arg\min} \: \mathbb{E}_{(\mathbf{v}, \mathbf{y}) \sim \mathcal{D}_{\text{fault}}} \left[ -\log p_{\theta_{\text{base}} \oplus \theta_{\text{LoRA}}}(\mathbf{y}|\mathbf{v}) \right]
\end{equation}
where $\theta_{\text{base}} \oplus \theta_{\text{LoRA}}$ denotes the composition of frozen base parameters with trainable LoRA parameters, enabling efficient adaptation while mitigating catastrophic forgetting of the model's generative capabilities.

A distinctive advantage of GFC is its seamless integration with follow-up analysis capabilities. By preserving the generative nature of the underlying LLM, AeroGPT can not only classify faults but also provide interpretable labels and respond to queries about fault characteristics. This is formalized as a conditional generation task:
\begin{equation}
p_{\theta}(\mathbf{r}|\mathbf{v}, \mathbf{y}, \mathbf{q}) = \prod_{t=1}^{|\mathbf{r}|} p_{\theta}(r_t|\mathbf{z}_{\text{audio}}, \mathbf{y}, \mathbf{q}, r_{<t})
\end{equation}
where $\mathbf{r}$ represents the generated response, $\mathbf{q}$ denotes a follow-up query, and $\mathbf{y}$ is the initially generated fault label. This process enables contextual analysis that considers both the original vibration signal and the diagnostic history, facilitating deeper exploration of fault characteristics and implications.

GFC offers several substantial advantages over traditional classification approaches. It eliminates post-processing steps such as argmax operations and label mapping, which streamlines the diagnostic pipeline. The natural language outputs enhance interpretability by being immediately comprehensible to maintenance personnel without requiring specialized knowledge of model architectures or output interpretations. Additionally, the generative nature of GFC enables straightforward extensibility to new fault types without requiring architectural modifications, as the system can adapt to the evolving aerospace industry simply by including examples of new faults in the training data. Experimental results demonstrate that GFC not only achieves superior classification accuracy compared to traditional approaches but also provides richer diagnostic information with practical utility for maintenance operations in aerospace applications.

\section{Experimental Validation}
\subsection{Experimental Setup}
To systematically evaluate the performance of AeroGPT, a series of experiments were conducted using two widely recognized bearing datasets: the aerospace bearing dataset from DIRG and the aero-engine bearing dataset from HIT. These datasets represent diverse operational conditions and fault characteristics relevant to aerospace applications, providing a robust foundation for validating our approach.

The experimental validation encompasses three primary parts: First, an ablation study to assess the individual contributions of each component within the AeroGPT framework, specifically examining the impact of the Foundation Model (FM), Vibration Signal Alignment (VSA), and Generative Fault Classification (GFC). Second, comparative evaluations against state-of-the-art deep learning models to benchmark AeroGPT's performance in terms of accuracy, precision, and F1-score. {Third, qualitative analyses to demonstrate AeroGPT's capability to generate interpretable, text-based diagnostic outputs that provide insights beyond mere fault classifications.}

\subsubsection{Evaluation Metrics}
To quantitatively assess AeroGPT's diagnostic performance, we employed standard classification metrics including accuracy, precision, and F1-score. These metrics provide complementary perspectives on the model's effectiveness in correctly identifying bearing faults across diverse operational conditions. Additionally, we evaluated the model's performance separately for non-defective and defective categories to assess its balanced capability across different fault conditions.

Beyond these quantitative metrics, the qualitative aspects of AeroGPT's outputs were also evaluated, particularly focusing on their interactivity, interpretability, and actionability. This dual evaluation approach provides a comprehensive assessment of both classification accuracy and practical utility in aerospace maintenance contexts.

\subsubsection{Implementation Details}
AeroGPT was implemented using the PyTorch framework, with the HuggingFace Transformers library utilized for model architecture and training utilities. All experiments were conducted on a single NVIDIA A10 GPU with {24 GB} of VRAM. 

{The foundation model used in the experiments was initialized with the weights of Qwen2-Audio \cite{chu2024qwen2audioa}, which comprises a pre-trained audio encoder with 124 million parameters and a large language model with 7 billion parameters.} {To prepare vibration signals for input to the audio encoder, several signal processing steps were applied. First, all vibration signals underwent resampling to standardize the sampling rate to 16 kHz, ensuring compatibility with the pre-trained audio encoder. Second, amplitude normalization was performed on each signal segment to eliminate variations in signal strength across different operating conditions and sensor configurations. Specifically, peak normalization was applied to linearly scale each segment's amplitude to a fixed range of [-1.0, 1.0], enabling the model to focus on structural patterns and waveform characteristics rather than absolute signal intensity. Third, the normalized signals were quantized to 16-bit Pulse-Code Modulation (PCM) format and saved as WAV (.wav) files, maintaining the standard audio format expected by the encoder while preserving signal fidelity.} {For the Vibration Signal Alignment (VSA) stage, a dataset of 36,053 vibration-text paired samples was utilized to bridge the domain gap between general audio knowledge and bearing-specific vibration patterns. The datasets used for classification were split into training and test sets, with a split ratio of 8:2.}

During fine-tuning process, the LoRA rank was set to 16, with a scaling factor of 32. The learning rate was initialized at $1 \times 10^{-5}$, and a linear warm-up schedule was employed for the first 5\% of total training steps to ensure stability. The batch size was set to 32, with gradient accumulation over 16 steps to maximize GPU utilization while maintaining memory efficiency.

\subsection{Case 1: DIRG Bearing Dataset}

\subsubsection{Dataset Description}

The primary dataset for experimental validation was obtained from the Dynamic and Identification Research Group (DIRG) at Politecnico di Torino \cite{daga2019politecnico}. This aerospace-focused dataset features high-speed aeronautical bearings operating at speeds up to 35,000~rpm, accurately representing aerospace application conditions. Their bearing test rig comprises a shaft supported by three roller bearings, with controlled fault conditions introduced to one of them. Vibration data was collected using triaxial IEPE accelerometers mounted at two strategic locations on the bearing supports. The dataset encompasses seven distinct bearing conditions: one healthy state (0A) and six damage states (1A-6A) with precisely controlled conical indentations (150, 250, and 450 $\mu$m) on either the inner ring or a single roller. Data was recorded across multiple operational scenarios with shaft speeds ranging from 6,000 to 30,000~rpm and radial loads varying from 0 to 1800~N, at a sampling frequency of 51,200 Hz. To simulate the harsh operational environment, 4 dB of Gaussian white noise was added to the original signals. A total of 22,134 training samples and 7,259 test samples were utilized.

\begin{table}[htbp]
\centering
\caption{Ablation study results demonstrating the contributions of each component in AeroGPT.}
\label{tab:ablation}
\resizebox{\linewidth}{!}{%
\begin{tabular}{ccc|ccc}
\toprule
 FM & VSA & GFC & Accuracy & Precision & F1-score \\
\midrule
$\checkmark$ & & & 14.87\% & 6.31\% & 4.20\% \\
$\checkmark$ & $\checkmark$ & & 20.65\% & 12.47\% & 10.83\% \\
$\checkmark$ & & $\checkmark$ & 97.21\% & 97.84\% & 97.52\% \\
$\checkmark$ & $\checkmark$ & $\checkmark$ & 98.94\% & 99.16\% & 99.02\% \\
\bottomrule
\end{tabular}
}
\end{table}

\subsubsection{Ablation Study}

\autoref{tab:ablation} presents the ablation study results for three components: Foundation Model (FM) with general audio understanding, Vibration Signal Alignment (VSA) for domain adaptation, and Generative Fault Classification (GFC) for direct label generation. {For FM+VSA, evaluation adopts a zero-shot protocol where the model selects from predefined fault categories, with top-1 accuracy computed based on exact matches.}

Using FM alone yields only 14.87\% accuracy, approximately equivalent to random guessing, underscoring the significant domain gap. Adding VSA improves accuracy to 20.65\%, indicating preliminary connections between audio knowledge and vibration patterns. However, FM+GFC achieves 97.21\% accuracy, demonstrating that task-specific fine-tuning enables effective adaptation even without explicit domain alignment.

The complete framework (FM+VSA+GFC) achieves the highest performance: 98.94\% accuracy, 99.16\% precision, and 99.02\% F1-score, corresponding to relative error reductions of 62.1\%, 61.1\%, and 60.5\% compared to FM+GFC. These results confirm the synergistic effect of VSA, which, while providing minimal benefits in isolation, significantly enhances GFC effectiveness by aligning representations with the aerospace bearing vibration domain.

\subsubsection{Comparison with Existing Methods}

\begin{table*}[htbp]
    \centering
    \caption{{Performance comparison of AeroGPT against both fault diagnosis models and general-purpose deep learning models on the DIRG dataset.}}
    \label{tab:comparison}
    \resizebox{\linewidth}{!}{%
    \begin{tabular}{@{}cccccccccc@{}}
    \toprule
    \multirow{2}{*}{Method} & \multicolumn{3}{c}{Non-Defective} & \multicolumn{3}{c}{Defective} & \multicolumn{3}{c}{Total} \\ 
    \cmidrule(l){2-4} \cmidrule(l){5-7} \cmidrule(l){8-10}
     & Accuracy & Precision & F1-score & Accuracy & Precision & F1-score & Accuracy & Precision & F1-score \\ 
    \midrule
    \multicolumn{10}{c}{Fault Diagnosis Models} \\ \midrule
    AeroGPT (Ours)                 & \textbf{99.46\%} & 97.08\%          & \textbf{98.14\%}          & \textbf{98.42\%}          & \textbf{99.26\%} & \textbf{99.07\%} & \textbf{98.94\%} & \textbf{99.16\%} & \textbf{99.02\%} \\
    {DCNDSC}                  & {99.20\%}          & {98.83\%}          & {97.21\%}          & {96.14\%}          & {97.48\%}          & {97.75\%}          & {97.67\%}          & {97.71\%}          & {97.68\%}          \\
    {LiConvFormer}            & {98.87\%}          & {97.08\%}          & {95.97\%}          & {93.57\%}          & {96.08\%}          & {96.26\%}          & {96.22\%}          & {96.24\%}          & {96.22\%}          \\
    {TFN-STFF}                & {96.90\%}          & {87.06\%}          & {89.46\%}          & {88.20\%}          & {93.53\%}          & {93.08\%}          & {92.55\%}          & {92.60\%}          & {92.55\%}          \\
    MA1DCNN                 & 99.15\%          & \textbf{99.18\%}          & 96.96\%          & 93.26\%          & 95.73\%          & 96.08\%          & 96.21\%          & 96.31\%          & 96.22\%          \\
        {ILDM}                    & {96.35\%}          & {84.10\%}          & {87.78\%}          & {76.07\%}          & {86.60\%}          & {85.93\%}          & {86.21\%}          & {86.37\%}          & {86.19\%}          \\
    {MPSO-ACBCNN}             & {99.01\%}          & {95.73\%}          & {96.56\%}          & {95.67\%}          & {97.61\%}          & {97.47\%}          & {97.34\%}          & {97.35\%}          & {97.34\%}          \\
    {TAUN}                    & {99.04\%}          & {98.21\%}          & {96.57\%}          & {93.50\%}          & {95.96\%}          & {96.22\%}          & {96.27\%}          & {96.30\%}          & {96.28\%}          \\
    {DSRSN}                   & {97.51\%}          & {86.77\%}          & {91.78\%}          & {91.56\%}          & {96.01\%}          & {95.02\%}          & {94.53\%}          & {94.77\%}          & {94.56\%}          \\
    {SPCFormer}               & {98.10\%}          & {93.68\%}          & {93.32\%}          & {84.90\%}          & {91.14\%}          & {91.20\%}          & {91.50\%}          & {91.61\%}          & {91.54\%}          \\
    {MRCNN-LSTM}              & {93.98\%}          & {78.63\%}          & {79.04\%}          & {68.30\%}          & {81.56\%}          & {81.49\%}          & {81.14\%}          & {81.25\%}          & {81.17\%}          \\

    \midrule
    \multicolumn{10}{c}{General Purpose Models} \\ \midrule
    ResNet50                & 98.14\%          & 93.42\%          & 93.47\%          & 87.35\%          & 92.63\%          & 92.63\%          & 92.75\%          & 92.97\%          & 92.78\%          \\
    ConvNeXt V2             & 95.55\%          & 84.53\%          & 84.40\%          & 72.05\%          & 83.68\%          & 83.70\%          & 83.80\%          & 83.77\%          & 83.76\%          \\
    Conv-Transformer        & 96.11\%          & 86.27\%          & 86.40\%          & 71.16\%          & 83.19\%          & 83.17\%          & 83.64\%          & 83.77\%          & 83.32\%          \\
    Masked AutoEncoder      & 98.29\%          & 94.62\%          & 93.98\%          & 84.96\%          & 91.13\%          & 91.23\%          & 91.62\%          & 91.61\%          & 91.61\%          \\
    PoolFormer              & 94.79\%          & 80.94\%          & 82.02\%          & 67.19\%          & 81.00\%          & 80.82\%          & 80.99\%          & 80.99\%          & 80.97\%          \\
    Swin-Transformer        & 95.91\%          & 87.83\%          & 85.26\%          & 79.27\%          & 87.55\%          & 87.96\%          & 87.59\%          & 87.82\%          & 87.65\%          \\
    Swin-Transformer V2     & 97.60\%          & 90.59\%          & 91.71\%          & 80.69\%          & 88.90\%          & 88.71\%          & 89.14\%          & 89.27\%          & 89.19\%          \\ 
    \bottomrule
    \end{tabular}
    }
\end{table*}

{To comprehensively assess AeroGPT's effectiveness, we benchmarked it against two categories of models: specialized fault diagnosis methods and general-purpose deep learning architectures. The fault diagnosis baselines include DCNDSC \cite{qin_large_2024}, a large-scale dense connectivity framework employing depthwise separable convolutions for multi-machine diagnostics; LiConvFormer \cite{yan_liconvformer_2024}, which combines separable multiscale convolutions with broadcast self-attention; TFN (with STFF) \cite{chen_tfn_2024}, a time-frequency network for bearing fault detection; MA1DCNN \cite{wang2020understanding}, a multi-scale attention convolutional network for vibration analysis; ILDM \cite{jia_industrial_2025}, a large-scale diagnostic model approach utilizing time-series embeddings with 1D-2D-1D transformations; DSRSN \cite{xu_deep_2024}, a deep spiking residual shrinkage network that introduces attention mechanisms and soft thresholding to improve recognition under high-noise conditions; TAUN \cite{lan_traceable_2025}, a traceable algorithm unrolling network that constructs an interpretable feature extractor by unrolling the iterative sparse coding algorithm; MPSO-ACBCNN \cite{zhao_automated_2024}, an automated CNN design method using modified particle swarm optimization with advanced convolution blocks for satellite attitude control system fault diagnosis; MRCNN-LSTM \cite{liu_model_2023}, a model fusion approach combining multiscale residual CNN with LSTM to extract both spatial and temporal features; and SPCFormer \cite{guo_effective_2025}, a lightweight Transformer with selective patches and channels modules. The general-purpose models comprise ResNet50 \cite{he2016deep}, ConvNeXt V2 \cite{woo_convnext_2023}, Conv-Transformer, Masked AutoEncoder \cite{he_masked_2021}, PoolFormer \cite{yu_metaformer_2022}, and Swin-Transformer variants \cite{liu_swin_2021,liu_swin_2022}. For vision-based architectures, 1D vibration signals were reshaped into approximately square 2D feature maps of size $d\times d$ where $d=\lfloor \sqrt{L}\rfloor$, with zero-padding applied when necessary.}

{As shown in \autoref{tab:comparison}, AeroGPT achieves superior performance across virtually all metrics, with overall accuracy reaching 98.94\%. Among fault diagnosis baselines, DCNDSC (97.67\%) and MPSO-ACBCNN (97.34\%) demonstrate strong performance through their specialized architectural designs, while TAUN (96.27\%) and LiConvFormer (96.22\%) also achieve competitive accuracy. In contrast, methods such as SPCFormer (91.50\%) and MRCNN-LSTM (81.14\%) show limited effectiveness on this dataset, likely due to their original design focus on different application domains. General-purpose models exhibit even larger performance gaps, with ResNet50 achieving only 92.75\% accuracy. These results demonstrate that domain-agnostic architectures, despite their sophistication, cannot adequately capture fault-specific patterns without specialized adaptation.}

{A critical observation concerns the performance disparity between defective and non-defective states. General-purpose models exhibit severe imbalance, with accuracy gaps exceeding 20 percentage points (e.g., ConvNeXt V2: 95.55\% vs. 72.05\%; PoolFormer: 94.79\% vs. 67.19\%). Even specialized models like TFN-STFF (96.90\% vs. 88.20\%) and ILDM (96.35\% vs. 76.07\%) show notable degradation on defective samples, indicating difficulties in discriminating among diverse fault types. Conversely, AeroGPT maintains exceptional balance with only a 1.04 percentage difference (99.46\% vs. 98.42\%), which is crucial for aerospace applications where both missed detections and false alarms carry substantial operational consequences.
}

{AeroGPT's advantages stem from two key factors. First, it leverages pre-trained acoustic representations that inherently capture temporal-spectral patterns relevant to mechanical vibrations, rather than learning from scratch. Second, the VSA adaptation stage effectively transfers this knowledge to aerospace-specific fault characteristics, as validated by ablation studies.}

\subsubsection{{Qualitative Analysis}}

\begin{figure*}[h]
    \centering
\includegraphics[width=0.95\linewidth]{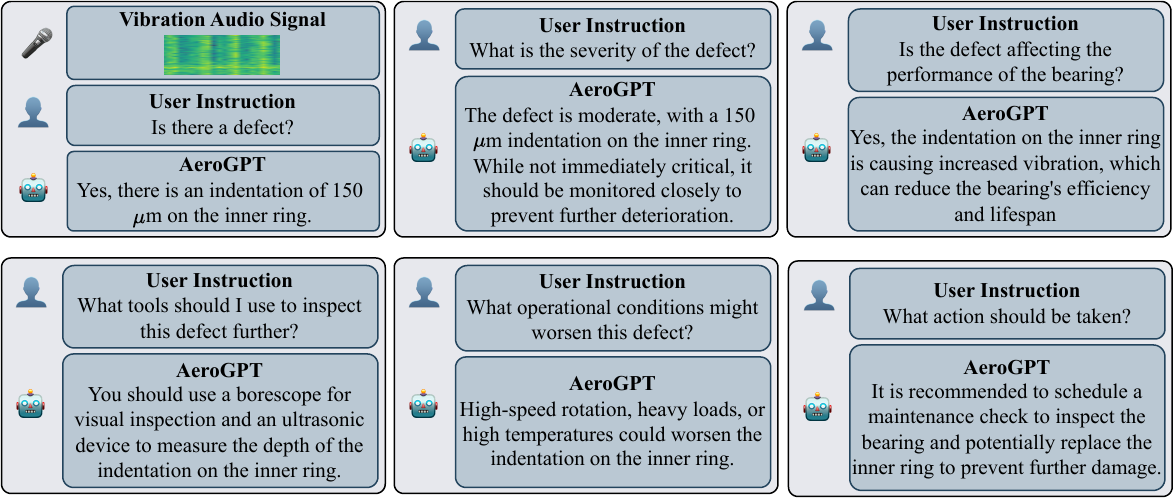}
    \caption{Examples of AeroGPT's generative fault diagnosis capability and answers to follow-up queries.}
    \label{fig:qualitative_case1}
\end{figure*}

{Beyond quantitative metrics, AeroGPT's interactive diagnostic capabilities were evaluated to assess practical utility in aerospace maintenance scenarios. \autoref{fig:qualitative_case1} demonstrates how AeroGPT transforms fault diagnosis from isolated classification to interactive consultation.}

{Instead of producing a simple fault label, AeroGPT generates detailed fault descriptions identifying the fault as ``inner ring indentation'' with a severity of ``150 $\mu$m.'' When questioned about severity, the system classifies the defect as ``moderate'' while noting it ``should be monitored closely to prevent further deterioration.'' Moreover, AeroGPT exhibits causal reasoning by articulating how inner ring indentation directly causes ``increased vibration'' and reduced ``efficiency and lifespan,'' providing mechanical insights that traditional classifiers cannot offer.}

{The framework further demonstrates domain expertise through contextually appropriate technical recommendations. When queried about inspection methodologies, AeroGPT recommends ``a borescope for visual inspection and an ultrasonic device to measure the depth of the indentation.'' When asked about exacerbating factors, it identifies ``high-speed rotation, heavy loads, or high temperatures'' as operational parameters that could accelerate deterioration. Additionally, the response prescribing a ``maintenance check to inspect the bearing and potentially replace the inner ring'' aligns with established aerospace protocols, effectively bridging the gap between fault detection and maintenance action.}

{The interactive dialogue capability demonstrates the advantage of the generative formulation: rather than outputting probability distributions requiring expert interpretation, AeroGPT directly produces human-readable diagnoses and engages in follow-up reasoning about fault mechanisms, severity assessment, and maintenance recommendations. This end-to-end interpretability eliminates the need to translate model outputs into maintenance actions. When combining with techniques such as RAG, AeroGPT can access a vast knowledge base to provide more accurate and contextually relevant responses.}

\subsection{Case 2: HIT Bearing Dataset}
\subsubsection{{Dataset Description}}

The second dataset was obtained by Hou \textit{et al.} from the Harbin Institute of Technology (HIT) \cite{hou2023intershaft}, featuring inter-shaft bearing data from an aero-engine system. Their experimental setup comprises a modified aero-engine with the critical dual-rotor structure preserved. The test rig follows a similar structure to that in the first case, {with the low-pressure rotor connected to the high-pressure rotor through a shaft and the inter-shaft bearings subjected to various fault conditions.} The system is equipped with two independent motors for the high and low pressure rotors, allowing for precise control over operational parameters. All signals were sampled at {25 kHz}, with tests conducted across 28 distinct operating conditions varying both speed ({ranging from 1000~rpm to 6000~rpm}) and speed ratio between the rotors (1.2 to 1.8). The dataset encompasses 2412 sets of vibration data labeled as three categories (healthy, inner ring defect, and outer ring defect), segmented into 20480-point sequences. The obtained training set consists of 1929 samples and the test set contains 483 samples.

\subsubsection{Comparison with Existing Methods}

\begin{table}[htbp]
    \centering
    \caption{{Performance comparison of AeroGPT against both fault diagnosis models and general-purpose deep learning models on the HIT dataset.}}
    \label{tab:comparison_case2}
    \resizebox{\linewidth}{!}{%
    \begin{tabular}{@{}cccc@{}}
    \toprule
    Methods             & Accuracy          & Precision         & F1-score          \\ 
    \midrule
    \multicolumn{4}{c}{Fault Diagnosis Models} \\ \midrule
    AeroGPT (Ours)             & \textbf{100.00\%} & \textbf{100.00\%} & \textbf{100.00\%} \\
    LiConvFormer        & 99.79\%           & 99.80\%           & 99.79\%           \\
    TFN                 & 98.58\%           & 98.58\%           & 98.58\%           \\
    DCNDSC              & 99.39\%           & 99.40\%           & 99.39\%           \\
    ILDM                & 98.76\%           & 98.77\%           & 98.76\%           \\
    MA1DCNN             & 99.59\%           & 99.59\%           & 99.59\%           \\
    {MPSO-ACBCNN}         & {99.79\%}           & {99.79\%}           & {99.79\%}           \\
    {TAUN}                & {99.38\%}           & {99.39\%}           & {99.38\%}           \\
    {DSRSN}               & {98.38\%}           & {98.39\%}           & {98.38\%}           \\
    {SPCFormer}           & {98.35\%}           & {98.37\%}           & {98.34\%}           \\
    {MRCNN-LSTM}          & {94.42\%}           & {94.48\%}           & {94.44\%}           \\
    \midrule
    \multicolumn{4}{c}{General Purpose Models} \\ \midrule
    ResNet50            & 99.79\%           & 99.79\%           & 99.79\%           \\
    ConvNeXt V2         & 94.41\%           & 95.71\%           & 95.94\%           \\
    Conv-Transformer    & 99.17\%           & 99.18\%           & 99.17\%           \\
    Masked AutoEncoder  & 98.34\%           & 98.47\%           & 98.34\%           \\
    PoolFormer          & 99.59\%           & 99.50\%           & 99.75\%           \\
    Swin-Transformer    & 99.80\%           & 99.75\%           & 99.88\%           \\
    Swin-Transformer V2 & 99.59\%           & 99.50\%           & 99.49\%           \\ 
    \bottomrule
    \end{tabular}
   }
\end{table}

Experimental results on the HIT inter-shaft bearing dataset demonstrate that AeroGPT achieves perfect classification performance with 100.00\% accuracy, precision, and F1-score across both defective and non-defective categories, as shown in \autoref{tab:comparison_case2}. Comparative models also performed exceptionally well on this dataset. While the performance gap between AeroGPT and traditional approaches is narrower on this dataset compared to DIRG, AeroGPT maintains its advantage by eliminating all misclassifications. This consistent superiority across both aerospace bearing datasets confirms the effectiveness of transferring audio domain knowledge to aero-engine bearing fault diagnosis through the proposed methodologies.

\subsubsection{{Qualitative Analysis}}

\begin{figure*}[htbp]
    \centering
    \includegraphics[width=0.95\linewidth]{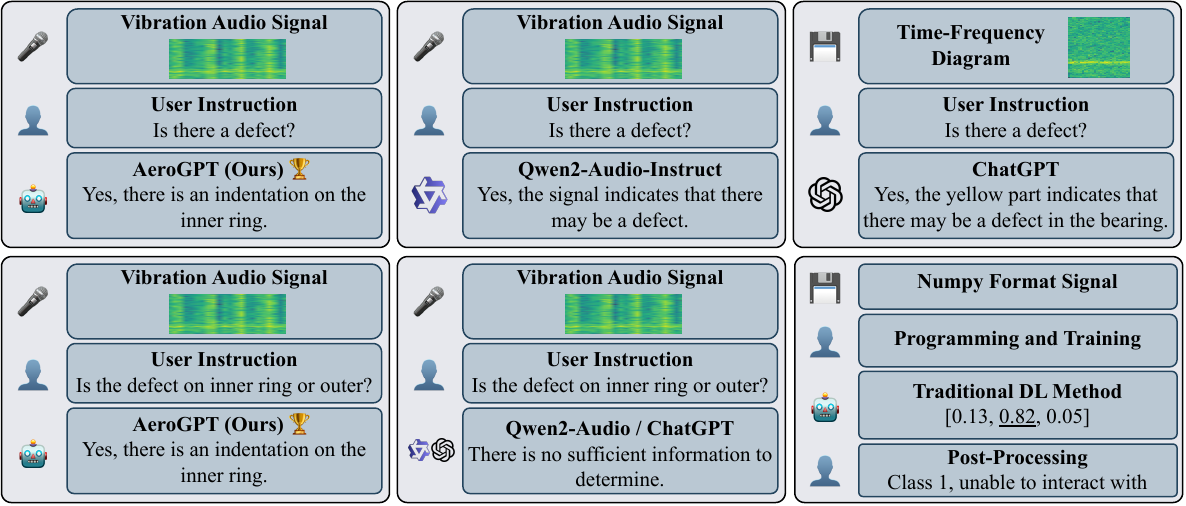}
    \caption{Comparison of AeroGPT's generative fault diagnosis ability with general-purpose models and conventional approaches.}
    \label{fig:qualitative_case2}
\end{figure*}

{The comparative qualitative evaluation illustrated in \autoref{fig:qualitative_case2} demonstrates AeroGPT's distinctive advantages over both general-purpose large language models and traditional deep learning approaches.}

{When presented with the same vibration signal, AeroGPT produces a definitive diagnosis stating ``Yes, there is an indentation on the inner ring,'' while general-purpose models can only make tentative assessments. Qwen2-Audio-Instruct produces the tentative ``Yes, the signal indicates that there may be a defect,'' and ChatGPT suggests ``Yes, the yellow part indicates that there may be a defect in the bearing.'' When asked to distinguish between inner and outer ring defects, AeroGPT confidently identifies the inner ring location, whereas both Qwen2-Audio and ChatGPT respond ``There is no sufficient information to determine.'' Traditional deep learning methods present different limitations: they output numerical probability distributions (e.g., [0.13, 0.82, 0.05]) requiring post-processing and expert interpretation, and remain inherently non-interactive.}

{This comparison highlights the core advantage of vibration-text alignment. While general-purpose audio models possess acoustic understanding capabilities, they lack domain-specific grounding for precise fault localization. The VSA stage bridges this gap by learning vibration-language correspondences, enabling AeroGPT to distinguish between fault locations that general models cannot differentiate. While conventional deep learning methods can also perform well on this task, they require post-processing and expert interpretation. These results validate AeroGPT's capabilities in achieving both high accuracy and interactivity in industrial fault diagnosis.}

{\section{Further Discussion}}

{\subsection{Computational Cost and Inference Latency}}

{For practical deployment considerations, we report the computational requirements and inference performance of AeroGPT. The foundation model comprises a pre-trained audio encoder with 124 million parameters and a large language model with 7 billion parameters. Using LoRA-based adaptation, only approximately 14 million parameters (0.2\% of the total) require training, significantly reducing computational requirements compared to full fine-tuning.}

{Under our experimental setup (NVIDIA A10 GPU with 24 GB VRAM, batch size 1), AeroGPT achieves the following performance characteristics: the average inference speed is 21.35 tokens per second, with an average time-to-first-token latency of 1.9 seconds, which means that simple fault diagnosis tasks can be completed within 5 seconds. The memory footprint during inference depends on the KV cache size, and a single GPU with 24 GB VRAM is sufficient for deployment. For training, the LoRA adaptation requires approximately 17 GB VRAM with gradient accumulation, enabling fine-tuning on a single GPU rather than requiring multi-GPU clusters.}

{These characteristics indicate that AeroGPT is suitable for offline diagnostic applications and near-real-time monitoring scenarios where latency requirements are on the order of seconds rather than milliseconds. For applications requiring faster response times, potential optimizations include model quantization, speculative decoding, or distillation to smaller architectures.}

{\subsection{Potential Misdiagnosis Risks}}

{For deployment in safety-critical aerospace applications, careful consideration of misdiagnosis risks is essential. Several potential failure modes must be considered. False negatives, i.e., missed faults, could lead to undetected bearing degradation and potential in-service failures; our experimental results show that AeroGPT achieves high recall (99.07\% on DIRG defective samples), but any missed detection in safety-critical applications warrants additional safeguards. False positives, i.e., false alarms, could trigger unnecessary maintenance actions, increasing operational costs; the high precision (99.26\% on DIRG defective samples) mitigates but does not eliminate this risk. Out-of-distribution (OOD) conditions, including novel fault types, operating regimes not seen during training, or sensor configurations different from training data, may produce unreliable predictions. Additionally, sensor faults such as corrupted, noisy, or missing input signals could lead to erroneous diagnoses.}

{\subsection{Safe Integration into Existing Workflows}}

{To safely and responsibly make use of AeroGPT in safety-critical aerospace applications, it is crucial to consider the potential risks and liabilities associated with its deployment. We emphasize that AeroGPT should be deployed as a decision-support module within an engine health monitoring workflow, not in the closed-loop engine control path. The framework is designed to augment rather than replace expert judgment in maintenance decision-making.}

{To mitigate misdiagnosis risks, we recommend several strategies for safe integration. First, an out-of-distribution (OOD) detection module should be placed upstream of the AeroGPT framework to filter out input signals that deviate significantly from the training distribution, such as those caused by sensor malfunctions, extreme operating conditions, or unknown interference, ensuring that the model is only queried with valid, high-fidelity data within its operational design domain. Second, a lightweight deterministic deep learning model can be deployed in parallel with AeroGPT as a cross-verification agent. If the generative output conflicts with the classification result of the deterministic model, the system should flag the discrepancy as a high-risk event for immediate investigation. Third, rule-based consistency checks against physics-based models, historical maintenance records, or redundant sensor readings can validate AeroGPT's outputs. Fourth, a human-in-the-loop review process should always be maintained for all critical maintenance decisions, with AeroGPT serving as a first-pass screening tool that prioritizes cases for expert attention. Fifth, online learning or periodic model updates should be implemented to adapt AeroGPT to emerging distribution shifts and novel fault patterns.}

{By positioning AeroGPT as an assistive diagnostic tool within a comprehensive health monitoring workflow that includes multiple redundant checks and human oversight, the framework can provide significant value in accelerating fault identification and reducing diagnostic workload while maintaining the safety standards required for aerospace applications.}

\section{Conclusion and Future Work}
This paper proposes AeroGPT, a novel framework based on a large-scale audio model that transfers knowledge from the general audio domain to aero-engine bearing fault diagnosis. By recognizing the intrinsic acoustic-like nature of bearing vibration signals, AeroGPT addresses fundamental limitations in current aero-engine fault diagnosis approaches. The framework's two key innovations, Vibration Signal Alignment (VSA) and Generative Fault Classification (GFC), provide a systematic methodology for adapting general audio knowledge to domain-specific vibration patterns while enabling direct generation of interpretable diagnostic outputs. Through comprehensive experimental validation on two aerospace bearing datasets, AeroGPT achieved exceptional performance with 98.94\% accuracy on the DIRG dataset and perfect classification accuracy on the HIT bearing dataset, surpassing traditional deep learning approaches. The qualitative analysis further demonstrated AeroGPT's unique capability to provide definitive, specific fault characterizations in natural language, contrasting sharply with the uncertainty exhibited by general-purpose language models and the post-processing requirements of conventional classification approaches. This transformation of fault diagnosis from complex analysis to intuitive conversation significantly enhances practical utility in aerospace maintenance contexts, where rapid, accurate interpretation is essential for preventing catastrophic failures. By eliminating the need for post-processing and providing interactive, interpretable, and actionable diagnostics, AeroGPT represents a significant advancement in aerospace reliability engineering, highlighting the substantial potential of large-scale audio models to revolutionize fault diagnosis in industrial settings. {Furthermore, we provide a detailed discussion on practical deployment considerations, including computational cost and inference latency, potential misdiagnosis risks, and strategies for safe integration into existing engine health monitoring workflows.}

{The promising performance of AeroGPT establishes a strong foundation for several exciting avenues of future research. A key direction for advancement is to broaden the scope of its operational validation. While the current study demonstrates efficacy on two aerospace engine datasets, future work could aim to assess the model's performance under a more extensive range of operating conditions, including extreme speeds, heavy loads, and transient states. This will provide a more comprehensive understanding of its reliability in real-world scenarios. Another important area for enhancement involves strengthening the model's generalization capabilities across diverse hardware and sensing configurations. Future investigations can focus on the model’s adaptability to variations in bearing types, sizes, and manufacturers, as well as different sensor placements and data acquisition systems, which can introduce significant data distribution shifts.}

\printbibliography

\end{document}